# Tunable Negative Differential Resistance controlled by Spin Blockade in Single Electron Transistors

M. Ciorga[1], M. Pioro-Ladriere[1,2], P. Zawadzki[1], P. Hawrylak[1], and A. S. Sachrajda[1],

[1]Institute for Microstructural Science, National Research Council of Canada,
Ottawa, K1A 0R6, Canada

[2]Département de Physique and CERPEMA, Université de Sherbrooke, Sherbrooke J1K 2R1, Canada

**Abstract**

We demonstrate a tunable negative differential resistance controlled by spin blockade in single electron transistors. The single electron transistors containing a few electrons and spin polarized source and drain contacts were formed in GaAs/GaAlAs heterojunctions using metallic gates. Coulomb blockade measurements performed as a function of applied source-drain bias, electron number, and magnetic field reveal well defined regimes where a decrease in the current is observed with increasing bias. We establish that the origin of the negative differential regime is the spin-polarized detection of electrons combined with a long spin relaxation time in the dot. These results indicate new functionalities that may be utilized in nano-spintronic devices in which the spin state is electro-statically controlled via the electron occupation number.

Over the last few years there has been a growing interest in semiconductor spintronics.[1] It is hoped that by utilizing the electron's spin degree of freedom devices could be designed with additional functionality and also that memory and logic functions could be combined on a single chip. The long spin coherence time has led to several proposals for spin based implementation schemes for quantum computers.[2] Here we focus on the effect of spin on the functions of a single electron transistor (SET). This is motivated by the current interest in nano-spintronics – the study and application of the spin properties of nanodevices. Spin effects in quantum dots such as the singlet-triplet transition, Hund's rule and the Kondo effect have been *indirectly* determined from addition spectra and temperature dependences using Coulomb blockade (CB) spectroscopic techniques (for a review and references see Ref. [3-7]). The effect of spin on CB has been investigated in quantum dots with unpolarized leads.[8-13] We have previously reported the observation of a different type of spin blockade (SB) resulting directly from the presence of spin polarized leads.[4,5] The effect manifested itself as a very strong amplitude modulation of CB peaks as a function of electron number or magnetic field. We made use of this SB spectroscopy to investigate *directly* spin transitions in quantum dots.[6,7] In this paper we present measurements made in the non-linear regime which reveal spin related negative differential resistance (NDR). We demonstrate that NDR regions observed in non-linear spectra are tunable and are related to our SB mechanism.

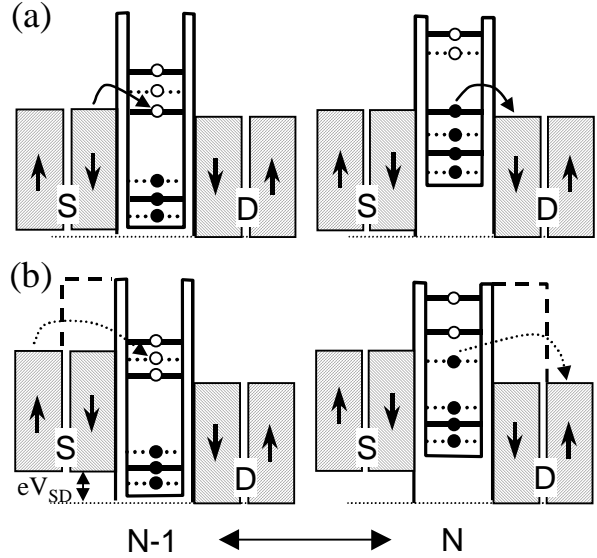

**Fig.1.** Schematic of the transport of Nth electron tunneling through the quantum dot containing N-1 electrons. a) not spin blockaded process in case of linear electronic transport b) SB NDR mechanism in non-linear electronic transport. Horizontal lines indicate occupied (●) and unoccupied (○) spin down (—) and spin up (⋯) levels; thick solid and thin dotted arrows show respectively high and low rate of the tunneling process; thick dashed line indicate thick barrier for spin up electrons entering/leaving the dot.

The principle of spin NDR is illustrated in Fig.1. Figure shows the schematic of our SET device. The source (S) and drain (D) leads consist of spatially separated spin up and down regions. The spin down (up) electrons tunnel from the source to the quantum dot and then to the drain via thin (thick) tunneling barriers. The effective thick barrier for the spin up electrons is due to the extra spatial separation resulting from the spin polarization at the 2DEG edge. The discrete quantum dot levels are assigned spin, depicted by horizontal solid (spin down) and dotted (spin up) lines. The charging energy gap existing between occupied (full circles) and unoccupied (open circles) levels in the dot is responsible for operation of SET: only one electron at the time can tunnel through the dot. Fig.1a. shows tunneling at low source-drain bias $V_{SD}$.

Only a spin down electron in the source can tunnel into the unoccupied spin down state of the dot. The spin down electron can now leave the dot to the spin down polarized drain, resulting in a high current, what is depicted in the picture with thick solid arrows showing a tunneling process. Fig.1b shows an additional tunneling process available at high $V_{SD}$. When $V_{SD}$ is high enough, tunneling through a spin-up state in the dot, in addition to spin-down tunneling, becomes possible. A spin up electron can enter the dot, through the thicker barrier (shown by dashed line), however, than spin-down electron in the previous case. If the spin relaxation time in the dot is sufficiently long, the spin up electron remains in the dot because of thick exit barrier for spin-up electrons and blocks the spin-down electron channel. Increasing of the bias voltage should then result in the decrease of current through the dot, i.e. in observation of NDR. The effect obviously depends on the arrangement of spin levels in the dot. This arrangement can be modified by adding an extra electron. Fig.2 shows the schematic of the same tunneling process for N+1 electron flowing through the dot containing already N electrons. Now at low $V_{SD}$ the current has to flow through a spin-up channel. Because it involves tunneling through thick barriers (shown by dashed lines) then the current is low (indicated by thin dotted arrows). Increasing $V_{SD}$ allows spin-down electrons to tunnel through a new spin-down level of the dot. This leads to increase of the current, and positive differential resistance (PDR). Therefore, to achieve NDR we need to realize spin polarized detection and a control over the spin levels of the dot through the electron number.

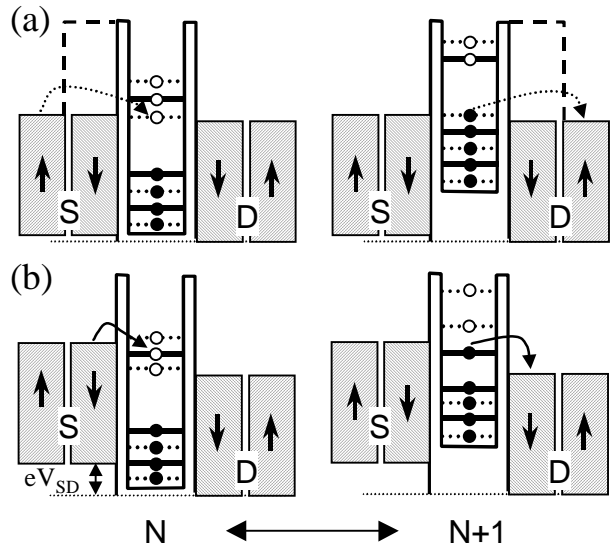

**Fig.2.** Schematic of the transport of N+1 electron tunneling through the quantum dot containing N electrons. a) spin blockaded process in case of linear electronic transport b) not blockaded PDR mechanism in non-linear electronic transport. The legend the same as in Fig.1.

The SEM picture of our experimental device is shown at Fig.3a. This device allows us to reduce the number of electrons confined in a lateral dot to zero. Further details of the device are given elsewhere. [5] Standard low power ac measurement techniques were used. An additional dc voltage $V_{SD}^{dc}$ in range of millivolts could also be applied. The measured signal is then proportional to the differential conductance $dI/dV_{SD}$. For the linear case, i.e. with $V_{SD}^{dc} = 0$, the standard CB peaks are reproduced. In this regime a peak in the current through the quantum dot is observed every time the electrochemical potential $\mu(N)$ of the dot matches the chemical potentials of the source ($\mu_S$) and the drain ($\mu_D$) leads. A non-linear regime is entered when a sufficient transport voltage $V_{SD}^{dc}$ is applied across the sample. When $V_{SD}$ becomes such that $eV_{SD} \equiv (\mu_S - \mu_D) > \Delta$, where $\Delta$ is a typical level spacing, the excited states of the quantum dot also contribute to the electronic current. It is important to note, however, that in the CB regime the current is restricted to flow one electron at a time so that an electron cannot enter the dot until an electron has left it. Adding a new channel with a finite occupation probability but a small exit probability therefore blocks the current ( the usual increase of current with an increase in the number of available states does not apply). This is seen as a negative peak in the $dI/dV_{SD}$ spectrum, and is an experimental observation of NDR behavior in quantum dots.

Fig.3b shows typical experimental data taken in the linear regime, illustrating the magnetic field evolution of a CB peak in the representative magnetic field range B=0.9-1.3T. The dark red (blue) color represents high (low) peak amplitude. The peak shows details of a very clear amplitude modulation. This type of modulation is observed only for B>0.5T and extends up to boundary of ν=2 phase in the dot. Based on calculated ground states of quantum dot in the vicinity of ν=2 boundary[7] this amplitude modulation can be related to spin effects occurring in quantum dots and explained in terms of spin injection/detection picture shown at figs. 1a, 2a. In our previous experiments on lateral dots it was shown that above a certain magnetic field (~ 0.5T) we were primarily injecting spin down polarized electrons into our lateral dot.[4,5] This occurs because the effective electron reservoirs in the system are the exchange enhanced spin polarized magnetic edge-states closest to the electrostatically defined tunneling barriers outside the dot. We stress that other amplitude modulation effects[14], not related to spin, also occur in specific regions of the phase diagram (e.g. the above large/small current regions are separated by small low current step regions related to a *spatial* tunneling amplitude effect within the dot) and are well understood and easily distinguished.

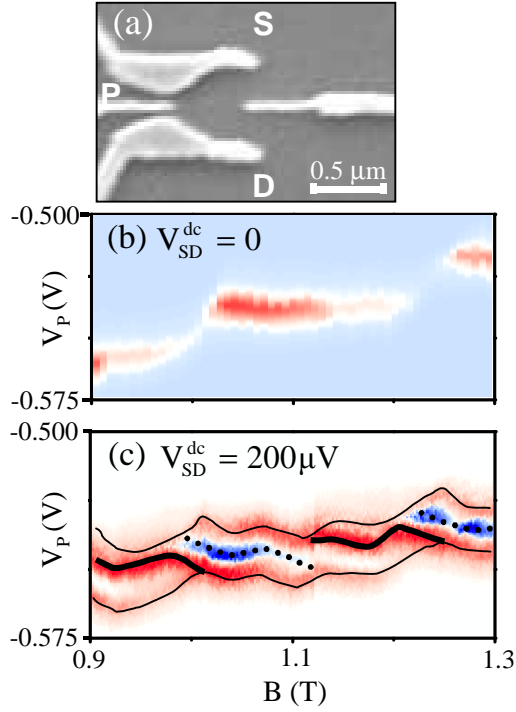

**Fig. 3.** a) SEM of single electron device; b) colored grayscale of $dI/dV_{SD}$ showing amplitude modulation of CB peak as a function of plunger voltage $V_P$ and magnetic field B from linear transport measurements; c) the same peak as in a) but in the non-linear regime. Dark blue color in the grayscale and dotted lines in the schematics indicate regions of negative differential resistance (NDR).

Figure 3c shows the same CB peak but in the non-linear regime. A source-drain voltage $V_{SD}^{dc} = 200 \mu V$ was applied to include the first excited state in the transport window $eV_{SD}$. Instead of a single peak current stripe is observed with a width $(\mu_S - \mu_D)/\alpha$, where $\alpha$ is the conversion factor between the plunger gate voltage $V_P$ and energy. The dark blue color on the grayscale represents the region where current decreased with increasing voltage i.e. the NDR region. The schematic drawing is superimposed on the grayscale to show more clearly the structure of the current stripe. The edge at the more (less) negative plunger voltage side represents the situation when $\mu(N) = \mu_S$ ($\mu_D$). Thick lines seen between both edges represent the evolution of the first excited state of N electron system, with dotted lines showing the NDR regions. We can clearly see that NDR regions are observed in the field range where the high amplitude of CB peak is measured in the linear regime. This behavior is in a very good agreement with spin polarised/injection detection picture from Fig.1. and confirms that the SB mechanism responsible for amplitude modulation in linear regime leads also to a NDR behavior in non-linear measurements.

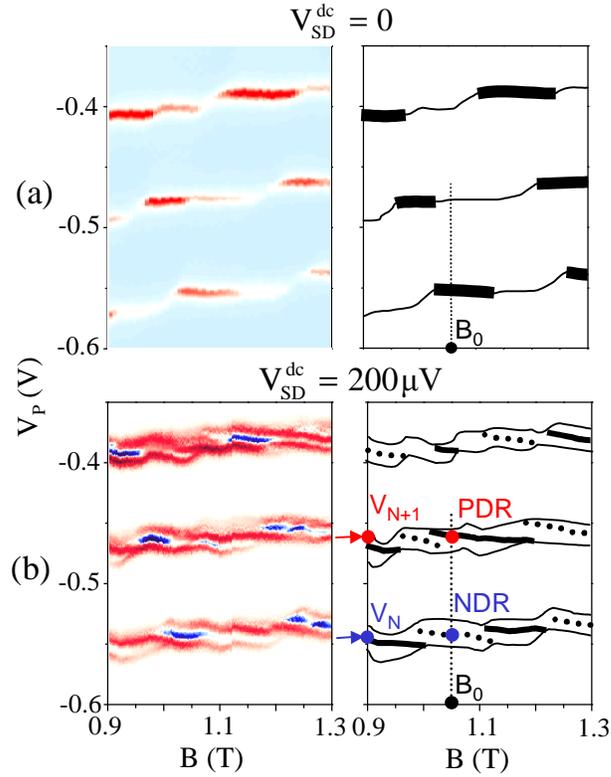

**Fig. 4.** Electron and magnetic field dependence of NDR and PDR. a) CB spectra from linear transport measurements. Left panel: colored grayscale of $dI/dV_{SD}$; Right panel: schematic representation of the same data with thick (thin) lines showing high (low) amplitude of CB peaks. b) the same peaks as in a) but in the non-linear regime. Dark blue color in the grayscale and dotted lines in the schematics indicate regions of negative differential resistance (NDR). Shown is the way of tuning NDR at constant magnetic field by changing the number of electrons in the dot by one.

In Fig. 4 we demonstrate that NDR is tunable with the number of electrons, that can be changed by means of plunger voltage $V_P$. The dependence of CB amplitude oscillations on occupation of the dot is shown in Fig. 4a. The figure shows typical experimental data taken in the linear regime, illustrating the magnetic field evolution of three CB peaks containing a very clear amplitude modulation. Figure 4b shows the same CB peaks but in the non-linear regime. The dark blue color on the grayscale represents the region where the NDR is observed. Thin solid lines in the accompanying schematic indicate the current stripe edges. We can clearly see that in all current stripes NDR region is observed in the field range where, for the same peak, the high amplitude of CB peak is measured in the linear regime. One can choose a certain fixed magnetic field value at which two subsequent CB peaks show an alternating behavior i.e. increasing the electron number from N to N+1 leads to opposite behavior of the CB amplitude: if for N electrons we have a peak in the amplitude, the amplitude for N+1 electrons is reduced. How this fact can be used to tune NDR with plunger voltage is shown on schematics which are accompanying grayscales. Let us choose two bottom CB peaks as an example. At magnetic field $B_0$ Nth electron CB peak shows behavior depicted in fig.1.: high amplitude in the linear regime and NDR behavior in the high source-drain measurements at plunger voltage $V_P=V_N$. Behavior of N+1 peak is the same as depicted at fig.2.: low current in linear regime and PDR behavior in non-linear regime. Changing $V_P$ from $V_N$ to $V_{N+1}$, at constant magnetic field $B_0$, switches then SET from NDR regime to PDR regime as number of electrons in quantum dot changes by one.

In summary, we have demonstrated the existence of NDR in a single electron transistor with spin polarized leads. NDR regions can be interpreted by means of spin dynamics and the dot spin states. Control over the number of electrons confined in the dot and over the spin of the dot allows one to tune the dot to switch between NDR and PDR regimes, i.e. these are in effect three terminal devices where the third terminal switches the sign of the differential resistance. We believe this to be a good 'proof of concept' example of the type of added functionality that spintronics will provide to circuit designers.[15]

**References**

1. S. Das Sarma, J. Fabian, X. Huand I. Žutič, Superlatt. and Microstr. **27**, 289 (2000); I. Malajovich, J.J. Berry, , Samarth, N. and D.D. Awschalom, Nature **411**, 770-772 (2001). Y. Ohno, , D.K. Young, , B. Beschoten, , F. Matsukura, H. Ohno and D.D. Awschalom, Nature **402**, 790 (1999); R. Fiederling, M. Keim, G. Reuscher, W. Ossau, G. Schmidt, A. Waag, A. and L.W. Molenkamp, Nature **402,** 787 (1999); P. R. Hammar, B. R. Bennet, M.J. Yang, and M. Johnson, Phys. Rev. Lett. **83**, 203 (1999).

2. D. Loss and D.P. di Vincenzo, Phys. Rev. A **57**, 120 (1998)

3. L.P. Kowenhoven, C.M. Marcus, P. McEuen, S. Tarucha, R. Westervelt and N.S. Wingreen, *Electron Transport in Quantum Dots* in *Mesoscopic Electron Transport*, edited by L. L. Sohn, L.P. Kouwenhoven and G. Schon, June 1996 (Kluwer, Series E 345, 1997)

4. P. Hawrylak, C. Gould, A.S. Sachrajda, Y. Feng, and Z. Wasilewski, Phys. Rev. B **59**, 2801 (1999)

5. M. Ciorga, A.S. Sachrajda, P. Hawrylak, C. Gould, P. Zawadzki, Y. Feng and Z. Wasilewski, Phys. Rev. B **61**, R16315 (2000).

6. M. Ciorga, A.S. Sachrajda, P. Hawrylak, C. Gould, P. Zawadzki, Y. Feng and Z. Wasilewski, Physica E **11,** 35 (2001).

7. A.S. Sachrajda, P. Hawrylak, M. Ciorga, C. Gould and P. Zawadzki, Physica E **10***,* 493 (2001).

8. T. Johnson, L.P. Kouwenhoven, W. de Jong, N.C. de Vaart, C.J.P.M. Harmans and C.T. Foxon, Phys. Rev. Lett. **69**, 1592 (1992)
9. J.T. Nicholls, J.E.F. Frost, M. Pepper, D.A. Ritchie, M.P. Grimshaw and G.A.C. Jones, Phys. Rev. B **48**, 8866 (1993)
10. J. Weis, R.J. Haug, K. v. Klitzing and K. Ploog, Phys. Rev. Lett. **71**, 4019 (1993)
11. D. Weinmann, W. Häusler and B. Kramer, Phys. Rev. Lett. **74**, 984 (1995); D. Weinmann, W. Haüsler, W. Pfaff and B. Kramer, Europhys. Lett. **26**, 467 (1994)
12. H. Imamura, H. Aoki and P. Maksym, Phys. Rev. B **57**, R4257 (1998)
13. H. Akera, Phys. Rev. B **60**, 10683 (1999)
14. P.L. McEuen, E.B. Foxman, U. Meirav, M.A. Kastner, N.S. Wingreen and S.J. Wind, Phys. Rev. Lett. **66**, 1926 (1991)
15. *Tunnel-Diode and Semiconductors Circuits*, ed. By J.M. Caroll (McGraw-Hill, New York, 1963)